\documentclass[twocolumn]{aastex631}
\usepackage{xspace,color,soul,lineno}

\newcommand{\per}{\ensuremath{^{-1}}\xspace}
\newcommand{\Lya}{Ly\ensuremath{\alpha}\xspace}

\shorttitle{Ultraluminous Ly$\alpha$ Luminosity Function Evolution}
\shortauthors{Taylor et al.}

\begin{document}

\title{The Evolution of the Ultraluminous Ly$\alpha$ Luminosity Function over
$z$=5.7--6.6}

\correspondingauthor{Anthony J. Taylor}
\email{ataylor@astro.wisc.edu}

\author[0000-0003-1282-7454]{A.~J.~Taylor}
\affiliation{Department of Astronomy, University of Wisconsin-Madison,
475 N. Charter Street, Madison, WI 53706, USA}

\author[0000-0002-6319-1575]{L.~L.~Cowie}
\affiliation{Institute for Astronomy, University of Hawaii,
2680 Woodlawn Drive, Honolulu, HI 96822, USA}

\author[0000-0002-3306-1606]{A.~J.~Barger}
\affiliation{Department of Astronomy, University of Wisconsin-Madison,
475 N. Charter Street, Madison, WI 53706, USA}
\affiliation{Department of Physics and Astronomy, University of Hawaii,
2505 Correa Road, Honolulu, HI 96822, USA}
\affiliation{Institute for Astronomy, University of Hawaii, 2680 Woodlawn Drive,
Honolulu, HI 96822, USA}

\author{E.~M.~Hu}
\affiliation{Institute for Astronomy, University of Hawaii,
2680 Woodlawn Drive, Honolulu, HI 96822, USA}

\author{A.~Songaila}
\affiliation{Institute for Astronomy, University of Hawaii,
2680 Woodlawn Drive, Honolulu, HI 96822, USA}

\begin{abstract}

Ultraluminous Lyman alpha (\Lya) emitting galaxies (ULLAEs) with $\log L (\textrm{\Lya})>43.5$~erg~s\per near the epoch of reionization $(z>5)$ make up the bright end of the LAE luminosity function (LF) and may provide insight into the process of reionization, including the formation of ionized bubbles around these extreme systems. We present a spectroscopic LF for ULLAEs at $z=5.7$. We used data from the HEROES
$\sim45$~deg$^2$ Subaru Hyper Suprime-Cam survey, which is centered on the North Ecliptic Pole 
and has both broadband $(grizY)$ and narrowband (NB816 and NB921) imaging, to select candidate 
ULLAEs based on a NB816 excess and a strong broadband Lyman break. 
We spectroscopically observed 17 ULLAE candidates with DEIMOS on Keck~II.
We confirmed 12 as LAEs at $z=5.7$, 9 of which are ULLAEs. 
The remaining sources are an AGN at $z=5.7$, an [OIII]$\lambda$5007 emitter at $z=0.63$, a red star, and two spectroscopic non-detections. 
Using the 9 confirmed ULLAEs, we construct a ULLAE LF at $z=5.7$. 
After applying a comprehensive incompleteness correction, 
we compare our new $z=5.7$ LF with our recent $z=6.6$ LF and with other LFs from the literature to look for evolution at the ultraluminous end. We find the overall ratio of the $z=5.7$ to $z=6.6$ ULLAE comoving number densities to be $1.92^{+1.12}_{-0.71}$, 
which corresponds to a LF offset of $0.28^{+0.20}_{-0.20}$~dex.

\end{abstract}

\keywords{Lyman-alpha, reionization, emission line galaxies, cosmology}

\section{Introduction}
In recent years, narrowband surveys and ground-based spectroscopic follow-up efforts have discovered a population of tens of ultraluminous ($\log L (\textrm{\Lya})>43.5$ erg s\per) Ly$\alpha$ emitting galaxies (ULLAEs) at redshifts z=6.6. These young and strongly star-forming galaxies may be the primary drivers of cosmic reionization, and they are unique probes of the intergalactic medium (IGM) near the epoch of reionization. The line shapes and luminosities of LAEs provide a powerful probe of the neutrality of
the IGM at these redshifts, and the ULLAEs at $z=6.6$ appear to mark the sites of giant ionized bubbles 
in the IGM \citep{hu16,matthee18,songaila18,taylor20,meyer21}. Such results
have been used to argue that the most luminous galaxies at these redshifts have high escape fractions and are producing the bulk of the ionizing photons \citep{naidu20}.
If correct, then this is a major paradigm shift from current models, which have the photons responsible for reionization arising from the very faint end of the galaxy luminosity function (LF).

While \Lya line profiles at $z\sim3$ show double peaks (both red and blue
peaks) in $\sim$30\% of cases \citep{kulas12}, at $z>5$, the blue peak is almost always scattered 
away by a neutral IGM \citep{hu10,hayes20}, leaving a single peak featuring a sharp blue break and an 
extended red wing. However, \cite{hu16} discovered a ULLAE 
in the COSMOS field at $z=6.6$ that had a blue peak (COLA1). 
Then \cite{songaila18} discovered another double-peaked ULLAE at $z=6.6$ 
in the HEROES survey in the North Ecliptic Pole (NEP) (source NELPA4).
More recently,
\cite{meyer21} found a third double-peaked emitter at $z=6.8$ at a sub-ultraluminous luminosity 
of $\log L (\textrm{\Lya})=42.99$ erg s\per in the A370p field (source A370p\_z1). 
In addition, \cite{bosman20} found the only double-peaked LAE at $z=5.8$ (source Aerith B), which has a sub-ultraluminous luminosity of 
$\log L (\textrm{\Lya})=43.03$ erg s\per and lies in a quasar proximity zone.
\cite{gronke20} showed that ionized bubbles around such objects can allow the
double-peaked structure to be seen.

Another way to explore the possibility of ionized bubbles is to determine the evolution of the ULLAE LF
over the redshift range $z=5.7$--6.6 from narrowband surveys.
\cite{santos16} were the first to claim no evolution in their photometric LAE LFs 
at the ultraluminous end after observing that their $z=5.7$ and $z=6.6$ LFs converged near 
$\log L (\textrm{\Lya})\approx 43.6$ erg s\per. They interpreted this result as evidence that the most luminous LAEs formed ionized bubbles around themselves, thereby becoming visible at earlier 
redshifts than less luminous LAEs. 

Spectroscopic confirmation of ULLAEs is essential to ensure sample fidelity and accurate LFs. Principal sources of contamination in ULLAE surveys include active galactic nuclei (AGNs) near the targeted \Lya redshift and lower redshift emission line galaxies, such as [OIII]$\lambda$5007 emitters. In addition, red stars can contaminate photometric narrowband samples. While survey depth is of primary importance for examining the faint end of the LAE LF, ULLAEs -- especially at high-redshift -- are very rare, requiring large but not overly deep surveys to achieve remotely significant number statistics (tens of sources). 
\cite{songaila18} and \cite{taylor20} analyzed the NB921 portion of the HEROES dataset to search for candidate ULLAEs at $z=6.6$,and \cite{taylor20} constructed the spectroscopic ULLAE LF at this redshift. 

In this paper, we analyze the NB816 portion of the HEROES dataset to search for candidate ULLAEs at $z=5.7$ and to construct the spectroscopic ULLAE LF at this redshift. We then test for evolution by comparing this $z=5.7$ ULLAE LF with its $z=6.6$ counterpart from \cite{taylor20}, as well as with LAE LFs from the literature.

We assume $\Omega_{\rm M}$=0.3, $\Omega_{\Lambda}$=0.7, and H$_0$=70~km~s\per Mpc\per throughout. We give all magnitudes in the AB magnitude system, where an AB magnitude is defined by $m_{AB} = -2.5 \log f_{\nu}-48.60$. We define $f_{\nu}$, the flux of the source, in units of erg~cm$^{-2}$~s\per~Hz\per.

\section{Candidate Selection}
\label{sec:sel}
We conducted our search for ULLAEs at $z=5.7$ using the \textul{H}awaii \textul{eRO}SITA \textul{E}cliptic Pole \textul{S}urvey (HEROES) data, which comprises $\sim$45 deg$^2$ of broadband $g$, $r$ (r2), $i$ (i2), $z$, $Y$, and narrowband NB816 and NB921 imaging from Hyper Suprime-Cam (HSC) on the Subaru 8.2~m telescope. \cite{songaila18} and \cite{taylor20} used the NB921 filter to search for ULLAEs at $z=6.6$. In repeating this search at $z=5.7$ using the NB816 filter, we followed a similar methodology for candidate identification and selection.

The $1\sigma$ noise in corrected (see Section \ref{sec:lum}) $2''$ diameter apertures in each broadband utilized in this study are $g$: 27.79, $r$: 27.07, $i$: 27.02, $z$: 26.66, and $Y$: 24.71. The NB816 depth is 25.70 with minor variations across the field. 
For more details on the observations and the data reduction for HEROES, see \cite{songaila18}. Briefly, all of the HEROES imaging was processed using the Pan-STARRS Image Processing Pipeline (IPP) \citep{magnier20a,magnier20b}. The original source detection was performed in the IPP. A source was added to a master catalog with Kron magnitudes measured in all 7 filters if it was detected at $5\sigma$ in any one of the 7 filters. The NB816 imaging was obtained on June 21, 29 and August 24, 26, 2017. The processed and stacked NB816 imaging has a median point spread function (PSF) full width half maximum (FWHM) of $0\farcs63$, varying from $0\farcs55$ to $0\farcs75$ across the field due primarily to variations in seeing conditions. The PSFs of the $i$ and $z$ bands are comparable, so color effects are not important with the fairly large apertures used.

\begin{figure}[ht]
\begin{centering}
\includegraphics[width=\linewidth]{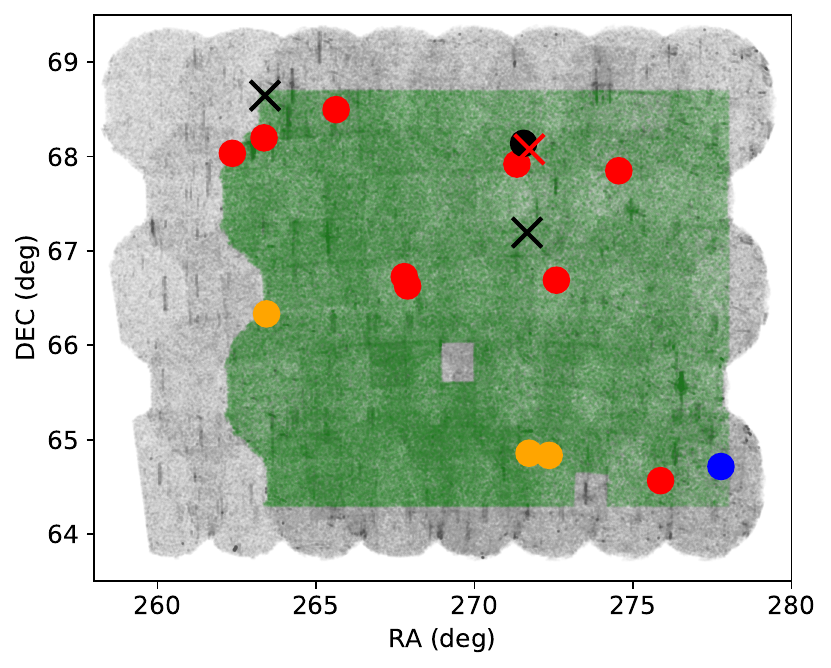}
\caption{
The gray shaded region shows the full $\sim$45 deg$^2$ of the current HEROES
survey, and the green shaded sub-region shows the most uniformly covered
26.48~deg$^2$ area targeted in this study. 
We mark the locations of the 9 spectroscopically confirmed $z=5.7$ ULLAEs
(red circles), the 3 spectroscopically confirmed $z=5.7$ sub-ULLAEs (orange circles), 
the $z=5.7$ AGN (black circle), the red star (red cross), the spectroscopic 
non-detections (black crosses), and the interloping $z=0.63$ $[\textrm{OIII}]\lambda5007$ 
emitter (blue circle). 
}
\label{fig:NEPmap}
\end{centering}
\end{figure}

We adopted the same $\sim30$~deg$^2$ area used by \cite{songaila18} (more precisely, 26.48~deg$^2$ due to the limited $r$ coverage; green shading in Figure~\ref{fig:NEPmap}). 
There are 149 individual images in this area, each with $11600\times11600$ pixels that cover $29'\times 29'$ with some minor overlap between images. 
This area contains 16,737,157 cataloged objects. 

We next applied several magnitude cuts to both the Kron and $2''$ aperture magnitudes for each source in the catalog. 
First, we required a significant brightness in NB816 ($18<\textrm{NB816}<23.5$). 
Second, we required non-detections $(>26)$ in $g$ and $r$ to identify strong Lyman breaks redward 
of the \Lya line. This is a relatively weak cut, since it corresponds to a $5\sigma$ level in $g$ and a $2.7\sigma$ level in $r$, on average.  
We did this deliberately to avoid preemptively rejecting candidate ULLAEs in areas of noisier data before a visual inspection.
Third, we required narrowband excesses relative to the $i$ $(i-\textrm{NB816}>1.3)$ and $z$ $(z-\textrm{NB816}>0.7)$ filters. These excesses select LAEs with observed-frame \Lya equivalent widths (EWs) of 130~\AA\ (rest-frame EWs of $\sim20$~\AA) or greater. Additionally, to help reject cosmic rays and data artifacts, we applied a $3\times3$ pixel median smoothing to the NB816 imaging and then remeasured the magnitudes, requiring NB816 (smoothed) $<24$. The median smoothing helps to reject hot pixels and cosmic rays, while only slightly dimming sources with Gaussian-like brightness profiles, hence the more relaxed 24th magnitude limit. After applying these cuts (which we summarize in Table~1), we had 995 $z=5.7$ ULLAE candidates. 

Finally, we visually inspected these candidates, rejecting those contaminated by glints, nearby bright stars, cosmic rays, data artifacts, etc. We further required no visible signal in the $g$ or $r$ bands. Most of the visually rejected candidates were faint $(g,r>26)$ but visible in the $g$ and $r$ bands, and were initially saved for visual inspection to account for any region-to-region variations in $g,r$ imaging depth. This resulted in a semi-final list of 24 candidates for spectroscopic follow-up. We ranked these candidates by estimated luminosity and visual quality, thereby producing our final list of 17 high-priority candidates.  The remaining 7 candidates were not bright enough in NB816 to be ultraluminous within our final targeted redshift range of $z=5.69-5.74$.

\begin{deluxetable}{cc}
\renewcommand\baselinestretch{1.0}
\tablewidth{0pt}
\tablecaption{Photometric Selection Criteria}
\scriptsize
\tablehead{Filter & Selection}
\startdata
$g$ & $>26$ \cr
$r$ & $>26$ \cr
NB816 & $>18$, $<23.5$\cr
NB816 (smoothed)& $<24$\cr
$i-\textrm{NB816}$ & $>1.3$ \cr
$z-\textrm{NB816}$ & $>0.7$ \cr
\enddata
\label{tab:selection}
\end{deluxetable}

\section{Spectroscopic Follow-Up}
\label{sec:spec}
Following \cite{songaila18} and \cite{taylor20}, we observed the 17 high-priority candidates with DEIMOS on Keck~II 
during an excellent run June 24--26, 2020. Briefly, we configured DEIMOS using the G830 grating and $1''$ slits, 
which provided a resolution of $R=2550$.  We took three 20-minute sub-exposures for each slitmask, dithering 
$\pm1\farcs5$ along the slit for each source for improved sky subtraction and minimization of CCD systematics. 
Each source received a total exposure time of 1 hour. 

We reduced the data using the standard pipeline from \cite{cowie96}. We performed an initial pixel-by-pixel sky subtraction by combining the three dithered exposures and subtracting the minimal value recorded by each pixel. Next, we median combined the three dithered frames, adjusting for the $\pm1\farcs5$ offsets. We rejected cosmic rays using a $3\times 3$ pixel median rejection spatial filter, and we quantified and corrected for geometric distortions in the spectra using pre-selected bright continuum sources from the slitmask. Lastly, we used the observed sky lines to calibrate the wavelength scale and to perform a final sky subtraction. 

Of the 17 observed high-priority targets, we identified 9 ULLAEs at $z=5.7$, 3 sub-ULLAEs at $z=5.7$, an AGN at $z=5.7$, a [OIII]$\lambda5007$ emitter at $z=0.63$, a red star, and 2 spectroscopic non-detections. We mark the locations of these sources in Figure~\ref{fig:NEPmap}. This corresponds to a 64\% (9/14) success rate in confirming $z=5.7$ ULLAEs (71\% (12/17) if the 
sub-ULLAEs are included), which is broadly consistent with our 64\% (9/14) success rate at confirming $z=6.6$ ULLAEs \citep{taylor20} and reflects a good balance between ensuring sample completeness and making efficient use of telescope time. This demonstrates the need for spectroscopic follow-up of high-redshift LAE candidates to reject low-redshift interlopers and red stars. 

We calculate our spectroscopic redshifts by taking the peak of the \Lya line in the one-dimensional (1D) spectra to be 1215.67 \AA{} in the rest-frame \citep[as in][]{songaila18,taylor20}. We note that the \Lya flux peaks may have velocity offsets relative to the galaxy systemic redshifts. However, the \Lya peak-determined redshifts are sufficient for calculating accurate transmission throughputs, and thus accurate line fluxes.  All 17 1D spectra are presented in Appendix \ref{append:a}. We will analyze the spectra in more detail in a future work
(A. Songaila et al.\ 2021, in preparation). 

After the DEIMOS run, when reviewing the candidate selection, we discovered one additional candidate (``NEP5.7CAND", in Table~\ref{tab:catalog}) that has not yet been spectroscopically confirmed. Based purely on our observed sample's ULLAE confirmation rate of 64\%, we include this source in our LFs (see Table~\ref{tab:lf}) with a weighting of 0.64 when counting the number of ULLAEs in our comoving volume.

\section{Line Fluxes and Luminosities}
\label{sec:lum}
We next calculated \Lya line fluxes and luminosities for our spectroscopically confirmed ULLAE sample. 
Since an absolute calibration for spectra without well-resolved continuum levels is notoriously difficult and unreliable, we instead opted to calculate \Lya fluxes from the narrowband imaging. In this method, we assume that all of the NB816 flux is due solely to the \Lya line. In reality, we expect a continuum contribution to this flux. However, any such contribution will be small (typically $<0.1$~dex), because the NB816 filter is narrow and the EW of the \Lya line is high.

We used  aperture-corrected $2''$ diameter aperture magnitudes, which we constructed by measuring the flux in both $2''$ and $4''$ diameter apertures centered on each of the spectroscopically confirmed ULLAEs. We then found the median difference between them for the sample (0.30~mags), which we subtracted from the raw $2''$ magnitudes to produce our final corrected $2''$ aperture magnitudes.

\begin{figure}[ht]
\includegraphics[width=\linewidth]{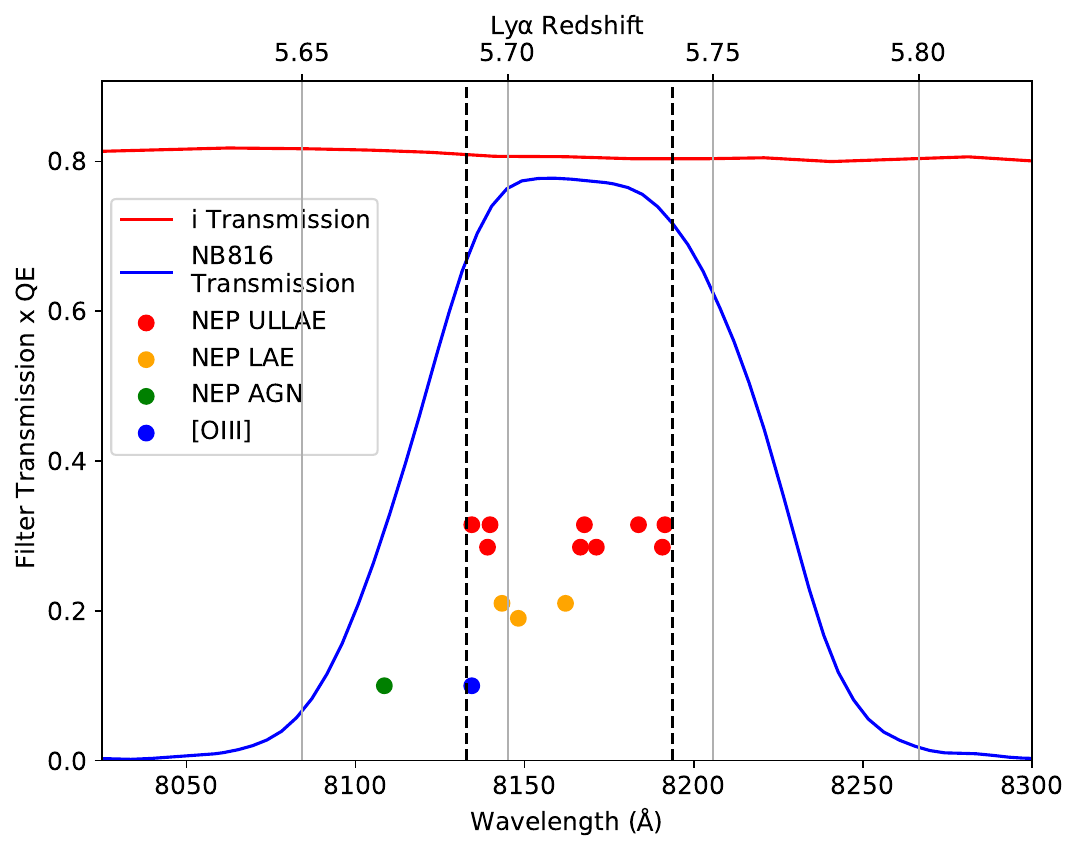}
\caption{Product of the filter transmission and CCD quantum efficiency for the 
HSC $i$ (red curve) and NB816 (blue curve) filters. Red circles show the 
redshift and observed-frame \Lya profile peak wavelength for each ULLAE. 
Orange circles show the same for the sub-ULLAEs. 
The green circle shows the $z=5.7$ AGN. 
The blue circle shows the weighted 
mean wavelength of the [OIII] doublet for the interloping [OIII] emitter. 
The vertical offsets between the colored circles are simply used to separate 
the different classes of objects and to avoid overlapping the circles. Black vertical dashed lines show the $z=5.69-5.74$ redshift bounds examined in this study.}
\label{fig:filters}
\end{figure}

After converting the NB816 corrected $2''$ diameter aperture magnitudes to fluxes, we divided the fluxes by the exact filter transmission efficiencies for their spectroscopic redshifts to calculate the full \Lya line fluxes. We then converted these \Lya line fluxes to \Lya line luminosities using redshift defined  cosmological luminosity distances. As a check, we compared the calculated narrowband line fluxes to the uncalibrated spectroscopic line fluxes and found a roughly linear correlation. In Figure~\ref{fig:filters}, we show the redshifts and the observed-frame \Lya (or [OIII]$\lambda5007$) wavelengths for the sample. In the same Figure, we also show the NB816 and $i$ filter transmission curves. Note that the sub-ULLAEs are fairly well centered at the peak of the NB816 transmission curve. If the \Lya observed wavelengths of these sub-ULLAEs had been more offset from the filter center, then they would have had higher luminosities and could potentially have been ULLAEs.  

In Table~\ref{tab:catalog}, we present our final catalog of 12 $z=5.7$ LAEs (9 ULLAEs), 
along with the $z=5.7$ AGN and the interloping [OIII] emitter.

\begin{deluxetable*}{cccccc}
\renewcommand\baselinestretch{1.0}
\tablewidth{0pt}
\tablecaption{Properties of the Spectroscopically Observed Sample}
\label{tab:catalog}
\tablehead{Source & R.A. & Decl. & NB816 & Redshift & $\log L($\Lya) \cr 
 & (deg) & (deg) & (AB) & & (erg~s$^{-1}$)}
\startdata
NEP5.7LA1 & 271.34009 & 67.920424 & 22.29 & 5.719 & 43.90 \\
NEP5.7LA2 & 267.89209 & 66.629250 & 22.55 & 5.696 & 43.81 \\
NEP5.7LA3 & 262.36445 & 68.034135 & 23.01 & 5.738 & 43.64 \\
NEP5.7LA4 & 272.58476 & 66.690398 & 22.88 & 5.738 & 43.69 \\
NEP5.7LA5 & 267.78810 & 66.727736 & 23.11 & 5.695 & 43.59 \\
NEP5.7LA6 & 274.55319 & 67.849627 & 23.12 & 5.722 & 43.57 \\
NEP5.7LA7 & 263.36466 & 68.198355 & 23.27 & 5.691 & 43.56 \\
NEP5.7LA8 & 265.63687 & 68.497512 & 23.28 & 5.718 & 43.50 \\
NEP5.7LA9 & 275.87088 & 64.565882 & 23.23 & 5.732 & 43.53 \\
NEP5.7LA10 & 263.44406 & 66.332895 & 23.36 & 5.714 & 43.47 \\
NEP5.7LA11 & 272.35809 & 64.833550 & 23.50 & 5.699 & 43.42 \\
NEP5.7LA12 & 271.72323 & 64.856770 & 23.55 & 5.703 & 43.39 \\
\hline
NEP5.7CAND & 271.01613 & 67.017966 & 23.21 & \nodata & $\sim$43.53 \\
\hline
NEP5.7AGN & 271.55017 & 68.140354 & 22.50 & 5.67 &  \nodata \\ 
\hline
OIII & 277.77502 & 64.715783 & 23.22 & 0.628 & \nodata \\ 
\enddata
\end{deluxetable*}

\section{Incompleteness Measurement}
To characterize the completeness of our ULLAE sample, we improved and refined the incompleteness simulation first presented in \cite{taylor20} using a new model spectrum, which we describe below. We generated artificial LAEs from this model spectrum, inserted them into our survey images, and then tried to recover and select them using the same criteria and methods that we used for our actual ULLAE photometric selection. 

We developed a simplified model rest-frame LAE spectrum based on the rest-frame median combination of our targeted LAE spectra. The model contained two components, the \Lya line and a flat continuum in frequency at wavelengths redward of the \Lya feature. We modeled the \Lya line as a simple right triangle with a peak at 1215~\AA{} and a linear decline to a red continuum flux at 1218~\AA{}. We defined the continuum flux level such that the \Lya line has a tunable rest-frame EW. 

We modified the model by adjusting its wavelength scale and amplitude to simulate LAEs at different redshifts and luminosities, respectively. 
We convolved each simulated LAE spectrum with the HSC $g$, $r$, $i$, $z$, and NB816 filter transmission curves to produce fluxes in each filter. We used these fluxes to scale the amplitude of a 2D Gaussian model with a FWHM of $0\farcs93$ at $z=5.7$ and $0\farcs73$ at $z=6.6$, which we then inserted into the survey imaging as an artificial LAE. We calibrated the FWHM of the 2D Gaussian model after stacking the images of the confirmed ULLAEs and confirming that the resulting stacked brightness profile was well fit by a 2D Gaussian model. At both redshifts, these profile FWHMs are larger than the seeing ($0\farcs93$ vs $0\farcs63$ at $z=5.7$ and $0\farcs73$ vs $0\farcs51$ at $z=6.6$); thus, the ULLAEs are partially resolved. In {\it HST} imaging, ULLAEs such as CR7 \citep{sobral19} can show complex structures and extended \Lya halos, but the seeing limits on our ground-based imaging smooth these structures into Gaussian brightness profiles. 

For all permutations of $z=5.66-5.78$ in 0.01 increment steps, $\log L$(\Lya)=42.5-44.0~erg~s$^{-1}$ in 0.05 increment steps, and rest-frame EW = 20, 25, 30, 40, 50, 60, 70~\AA{}, we scattered 1000 simulated LAEs into each image 
in the 5 relevant filters. We next used \textit{sep} \citep{sep}, a Python wrapper for \textit{SExtractor} \citep{sextractor}, to detect sources in the NB816 filter with a 5$\sigma$ detection threshold and to measure magnitudes for the detected NB816 source positions in all five filters. We then applied our photometric cuts (see Table~1) to the resulting source catalog and determined what fraction of the simulated sources we recovered and selected.  We took this fraction as our completeness value for the combination of the given image, redshift, and \Lya luminosity. 

\begin{figure}[h]
\includegraphics[width=\linewidth]{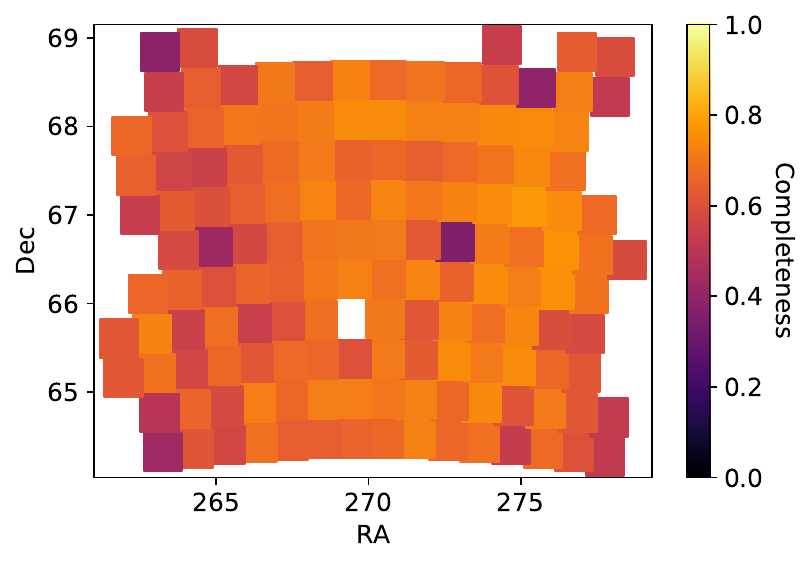}
\caption{Completeness measures integrated and averaged over $\log L(\rm \Lya$)=43.5--44.0~erg~s$^{-1}$, EW = 20--70~\AA{}, and $z=5.69$--5.74 for each of the 149 images. The center ``hole" in the data is a pointing that does not have $r$ filter coverage.
}
\label{fig:incskycells}
\end{figure}

\begin{figure}[h]
\includegraphics[width=\linewidth]{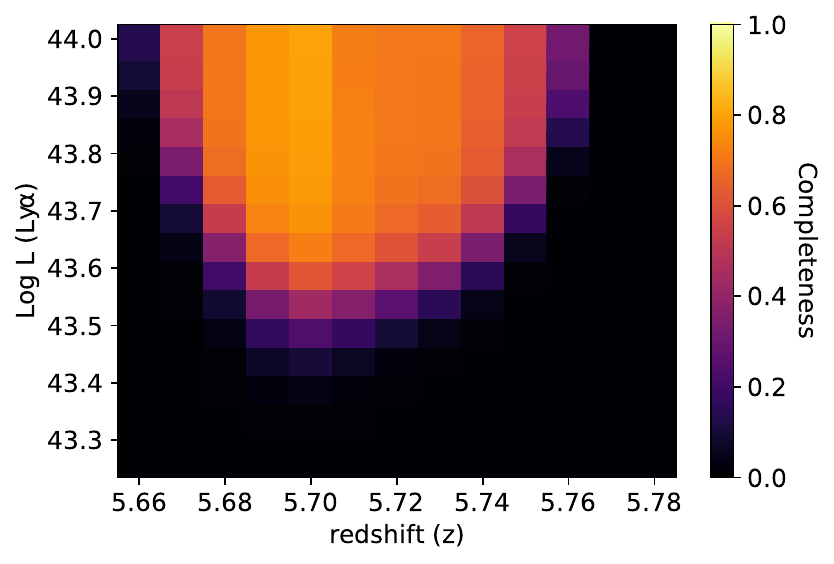}
\caption{Completeness measures averaged over all 149 images and EW = 20--70~\AA{}. 
This informed our chosen range of $z=5.69$--5.74 for the LFs.
}
\label{fig:incllyaz}
\end{figure}

We summarize the results of this analysis in Figures~\ref{fig:incskycells} and~\ref{fig:incllyaz}.
In Figure~\ref{fig:incskycells}, we show for each image in the survey field the completeness of ULLAEs ($\log L(\rm \Lya)=43.5--44.0$~erg~s$^{-1}$) averaged over EW = 20--70~\AA{} and our chosen redshift range of $z=5.69-5.74$ 
(see below). 
The overall average ULLAE completeness is 65\% across the field (77\% for $\log L(\rm \Lya$)=43.75--44.00~erg~s$^{-1}$, and 53\% for $\log L(\rm \Lya$)=43.50--43.75~erg~s$^{-1}$). This is very uniform across the survey images, with an image-to-image completeness standard deviation of 8\%. 

In Figure~\ref{fig:incllyaz}, we plot the $z=5.7$ survey completeness as a function of redshift and \Lya luminosity, averaging over the images in the field. The completeness roughly traces out the NB816 filter transmission profile in redshift space. This effect is primarily due to the NB816$<23.5$ magnitude cut from the photometric selection. We used these results to choose a redshift range of $z=5.69$--5.74 for the LF, since that range produces a sample with high expected completeness. 

For consistency, we applied our updated simulation procedures to the NB921 dataset from \cite{taylor20}. This resulted in some change in the incompleteness correction for our $z=6.6$ LF, especially at the faint end, due primarily to the new calculation over a range of EWs, the improved input spectrum, and the modified simulated source {brightness profile}. Here the overall average ULLAE completeness for our targeted redshift range of $z=6.52-6.62$ is 71\% across the field (80\% for $\log L(\rm \Lya$)=43.75--44.00~erg~s$^{-1}$, and 61\% for $\log L(\rm \Lya$)=43.50--43.75~erg~s$^{-1}$). The $\sim$6 percentage point difference in completeness between the $z=5.7$ and $z=6.6$ ULLAE samples is due primarily to the stringency of the NB816$-z$ cut at $z=5.7$, which does not have an analog in the $z=6.6$ selection, as well as the larger brightness profile at $z=5.7$, which reduces the detection rate of fainter sources.

\section{Luminosity Functions}
\label{sec:LFs}
In constructing the LFs for the $z=5.7$ and $z=6.6$ ULLAE samples, we first computed the comoving volumes for each redshift range. The 26.48 deg$^2$ NEP field from $z=5.69$--5.74 corresponds to a comoving volume of $1.19\times 10^7$~Mpc$^3$. The same field at $z=6.52$--6.62 encompasses $2.17\times 10^7$~Mpc$^3$, while the 3 deg$^2$ COSMOS field \citep{hu16} encompasses $2.46\times 10^6$~Mpc$^3$, for a total $z=6.6$ comoving volume of $2.42\times 10^7$~Mpc$^3$. 

We next define two bins in $\log L \rm(\Lya)$ space (43.5--43.75~erg~s$^{-1}$ and 43.75--44.0~erg~s$^{-1}$) in order to best compensate for the low number statistics of our 9 ULLAE sample at $z=5.7$. Despite confirming 3 sub-ULLAEs with $\log L \rm(\Lya) = 43.25$--43.5~erg~s$^{-1}$, the completeness of such a potential third bin (4\%) is too low to be considered in this study. Moreover, since sub-ULLAEs were never intentionally targeted in this study, this bin is not uniformly populated (e.g., the faintest LAE in our sample has a luminosity of $\log L \rm(\Lya) = 43.39$~erg~s$^{-1}$).

For each of the two redshift ranges, we calculated the number densities by dividing the number of ULLAEs in each bin by the total comoving volume for that redshift range.
We applied our incompleteness correction by dividing each uncorrected number density by its respective completeness. As our ULLAE sample is purely spectroscopic, we do not need to correct for low-redshift interlopers. Additionally, we do not need to correct for the non-rectangular shapes of the filter profiles, since with a spectroscopic sample, we know the precise filter transmission efficiencies for the redshifted \Lya lines.
With no other sources of contamination, our errors are based on low number statistics, and we calculate them using the methodology from \cite{gehrels86}. We give the completeness and the corrected and uncorrected number densities for both redshift ranges in Table~\ref{tab:lf}.

\begin{deluxetable*}{cccccc}[hbt]
\renewcommand\baselinestretch{1.0}
\tablewidth{0pt}
\tablecaption{Luminosity Function Data}
\label{tab:lf}
\tablehead{Redshift $z$ & $\log L(\textrm{\Lya})$ & Number of ULLAEs & Uncorrected $\log \phi$ & Corrected $\log \phi$ & Completeness  \cr
& $\Delta \log L (\textrm{\Lya})=0.25$ & & $\Delta \log L(\textrm{\Lya})^{-1}$ Mpc$^{-3}$ & $\Delta \log L(\textrm{\Lya})^{-1}$ Mpc$^{-3}$ & 
}
\startdata
$5.69-5.74$ & $43.50-43.75$ & 7.64 & $-5.591^{+0.149}_{-0.238}$ & $-5.313^{+0.149}_{-0.238}$ & 0.527  \cr
$5.69-5.74$ & $43.75-44.00$ & 2 & $-6.173^{+0.365}_{-0.451}$ & $-6.060^{+0.365}_{-0.451}$ & 0.771  \cr
$6.52-6.62$ & $43.50-43.75$ & 8 & $-5.878^{+0.174}_{-0.184}$ & $-5.665^{+0.174}_{-0.184}$ & 0.613  \cr
$6.52-6.62$ & $43.75-44.00$ & 3 & $-6.304^{+0.295}_{-0.341}$ & $-6.206^{+0.295}_{-0.341}$ & 0.799  \cr
\enddata
\end{deluxetable*}

\begin{figure*}[ht]
\includegraphics[width=\linewidth]{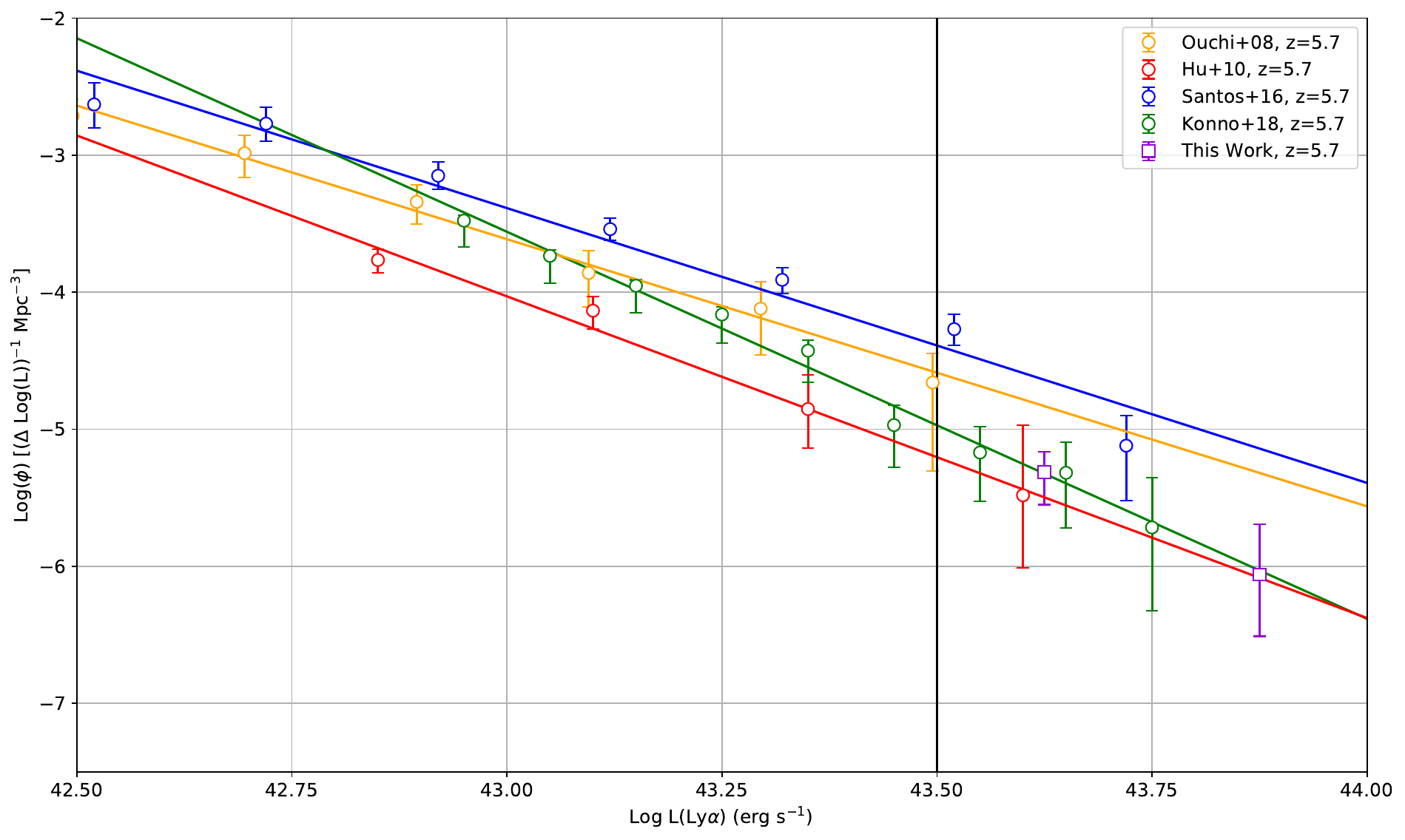}
\includegraphics[width=\linewidth]{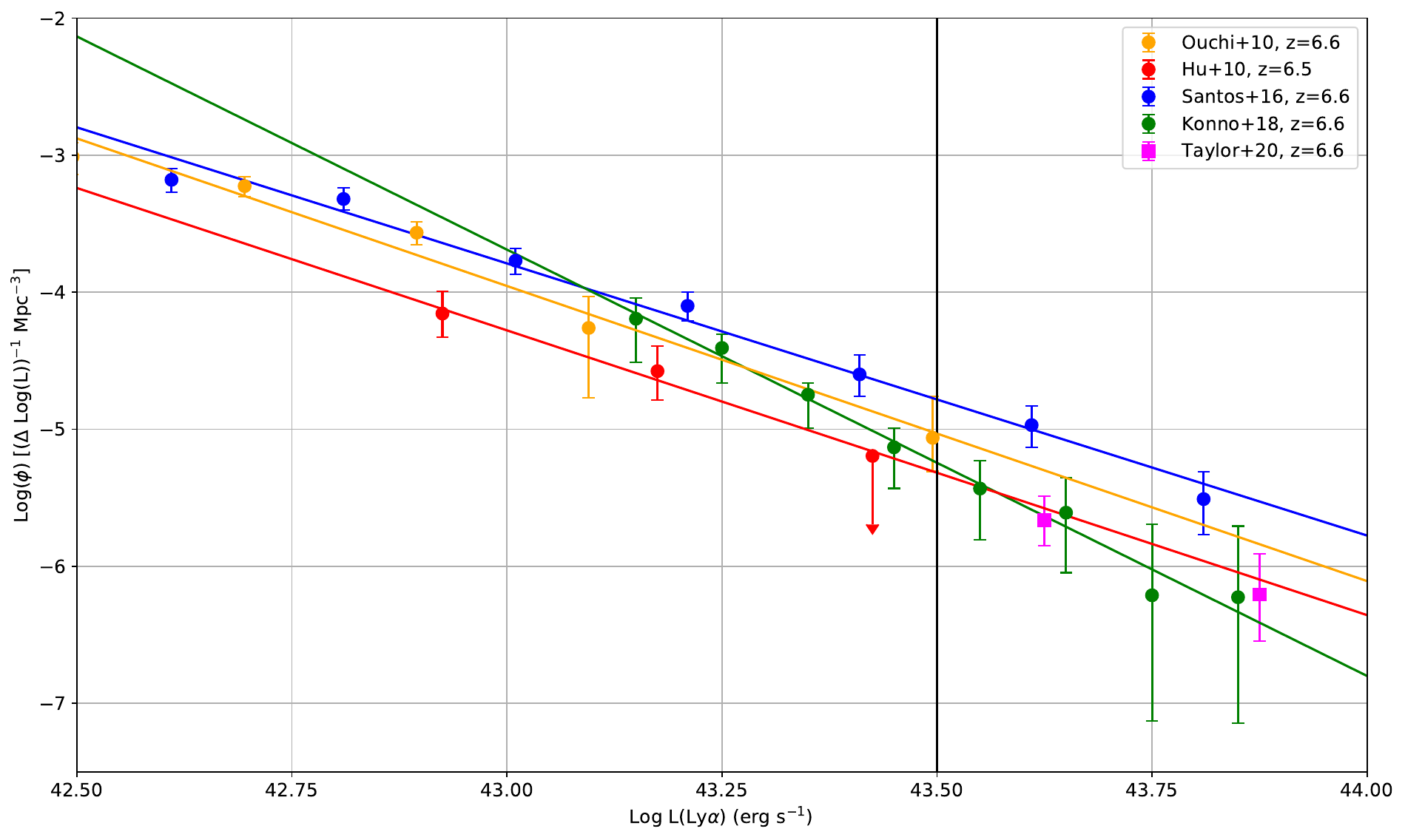}
\caption{
LF measurements for (a) our $z=5.7$ ULLAE sample  (purple squares) and (b) the ULLAE sample from \cite{taylor20} (pink squares). The bold vertical line defines our ultraluminous cutoff at $\log L(\rm \Lya)> 43.5$~erg~s$^{-1}$. For comparison, we plot literature LAE LFs at (a) $z=5.7$ \citep{ouchi10,hu10,santos16,konno18} and (b) $z=6.6$ \citep{ouchi08,hu10,santos16,konno18} (see legends for colors). Note that the $z=6.6$ LF of \cite{santos16} is an updated version from \cite{matthee15}. 
Power law fits to each literature study are shown in matching colored lines.
}
\label{fig:66LF}
\end{figure*}

In Figure~\ref{fig:66LF}(a), we show our ULLAE LF at $z=5.7$ (purple squares), and in Figure~\ref{fig:66LF}(b), we show the reprocessed \cite{taylor20} ULLAE LF at $z=6.6$ (pink squares). We supplement these with various LAE LFs from the literature for comparison. The matching colored lines for each literature sample are 

power law fits.
Above the ultraluminous threshold (shown as the bold vertical line at $\log L(\rm \Lya)=43.5$~erg~s$^{-1}$), our samples have smaller errors when compared to other samples, owing primarily to our comparatively large samples (9 ULLAEs at $z=5.7$ and 11 at $z=6.6$) drawn from our large survey area (26.48~deg$^2$).

In \cite{taylor20}, we highlighted the need for 
consistency in survey construction when comparing LFs across both redshift and luminosity. 
While nearly all of the literature LFs at either redshift agree with each other within their error bars, in order to clearly demonstrate evolution (or lack thereof) at the ultraluminous end of the LF, each study may only be compared to itself across redshifts.

Fortunately, the LFs from \cite{hu10} at both redshifts were constructed from purely spectroscopic samples based on Suprime-Cam (HSC's predecessor) photometry and DEIMOS spectroscopy. Thus, our spectroscopic LFs can largely be considered as ultraluminous extensions of their LFs, as both studies used consistent methodology.
At $z=5.7$, there is a slight overlap between the two LFs at $\log L(\rm \Lya)\sim 43.6$~erg~s$^{-1}$, and our value is well within the $1\sigma$ error bars of the \cite{hu10} value (see Figure~\ref{fig:66LF}(b)). 
(While it has not yet been made into a LF, we look forward to comparing with the recent 260 source $z=5.7$ LAE spectroscopic sample from \citealt{ning20} in the future.) The opacity of the IGM and the resulting \Lya transmission fraction changes rapidly at high redshifts \citep[e.g.][]{songaila04,laursen11}. Thus, the expectation would be that the number density of LAEs at $z=6.6$ would be lower than that at $z=5.7$. However, this may not be the case for the most luminous LAEs, if they are able to ionize the IGM around them \citep{santos16}.

\begin{figure*}[htb]
\includegraphics[angle=0,width=\linewidth]{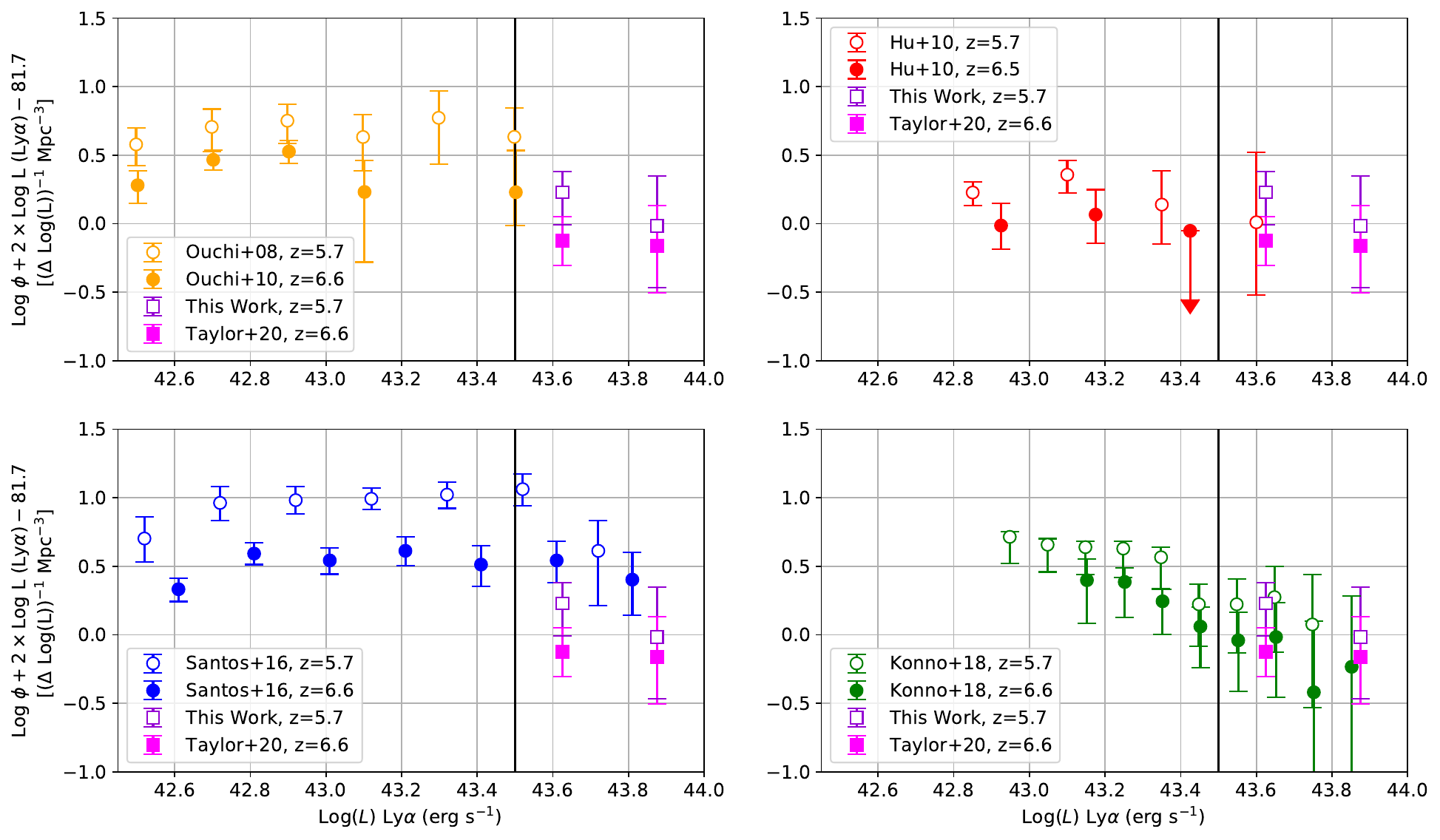}
\caption{Evolution of the \Lya LF over $z=5.7$--6.6 for the (a) \cite{ouchi08,ouchi10}, (b) \cite{hu10}, (c) \cite{santos16}, and
(d) \cite{konno18} surveys, shown relative to our $z=5.7$ ULLAE LF (purple squares) and
the reprocessed \cite{taylor20} $z=6.6$ ULLAE LF (pink squares).
In each panel, we subtract a power law with a fixed index of -2 and a normalization of 
81.707 $[\Delta \log L(\textrm{\Lya})]^{-1}$ Mpc$^{-3}$. Minor shifts in luminosity of $\pm0.002$ have been applied as needed to distinguish overlapping points and error bars.}
\label{fig:5766}
\end{figure*}

A number of papers have suggested that there is evolution of the LF over $z=5.7$--6.6 for sub-ULLAEs 
\citep{ouchi10,hu10,santos16}.
To illustrate this evolution, in Figure \ref{fig:5766}, we show the LFs from \cite{ouchi08,ouchi10}, \cite{hu10}, \cite{santos16}, and
\cite{konno18}, along with our ULLAE LFs. In log-log space,
we subtract a power law with a fixed index of -2 and a normalization of 81.707 $[\Delta \log L(\textrm{\Lya})]^{-1}$ Mpc$^{-3}$ from each LF in all panels.
We derived this power law from the best fit (with a fixed index of -2) to the \cite{hu10} $z=6.6$ data. 
As can be seen from Figure 6, this provides a good fit to most of the data sets, as described in \citet{taylor20}.
After subtracting this power law from each LF, it is easier to inspect the LFs for evolution both above and below the ultraluminous threshold.


In Figure~\ref{fig:5766}(a), we show the $\sim0.3$~dex offset at $\log L(\rm \Lya)< 43.5$~erg~s$^{-1}$ between the $z=5.7$ and $z=6.6$  LFs of \cite{ouchi08,ouchi10}, consistent with their claim of significant evolution from $z=6.6$ to $z=5.7$. However, as these LFs do not extend to ultraluminous luminosities, they are not directly comparable to our data.
In Figure~\ref{fig:5766}(b), we show the spectroscopic LFs of \cite{hu10}.  As previously stated, the pure spectroscopic nature of the \cite{hu10} LFs makes them directly comparable to our own, and we see good agreement within their error bars at both redshifts at the ultraluminous end.

{The photometric LFs from \cite{santos16} in Figure~\ref{fig:5766}(c) illustrate their claims of evolution (i.e., a $\sim0.4$~dex offset) at $\log L(\rm \Lya)< 43.5$~erg~s\per and convergence at
$\log L(\rm \Lya) > 43.6$~erg~s\per.  However, the $\sim$0.5~dex normalization offset of the \cite{santos16} LFs at both redshifts when compared to the other studies is concerning. It is important to stress how critical spectroscopic confirmation is for making a convincing case for the convergence of the LFs at the ultraluminous end. The offset of the \cite{santos16} LF by $\sim0.5-0.7$~dex from the \cite{hu10} LF is very likely attributable to the lack of spectroscopy for the \cite{santos16} sample, since their sample will inevitably suffer from a degree of contamination from lower redshift interlopers and/or an overestimation of incompleteness. Note that \cite{santos16} also have a $\sim0.5$~dex higher number density of ULLAEs than the we do.

In Figure~\ref{fig:5766}(d), the LFs from \cite{konno18} show agreement with the results of this study at the ultraluminous end. However, their very large error bars at the ultraluminous end and the markedly different slopes of their LFs when compared to the other studies make it difficult to draw confident conclusions from this dataset.

For our dataset, we find the overall ratio of the $z=5.7$ to $z=6.6$ ULLAE comoving number densities to be $1.92^{+1.12}_{-0.71}$.
When separated into two bins, we find the ratios to be $2.25^{+1.58}_{-0.93}$ for the
$\log L(\rm \Lya)=43.50-43.75$~erg~s\per bin and
$1.40^{+2.80}_{-0.98}$ for the $\log L(\rm \Lya)=43.75-44.00$~erg~s\per bin.
These values correspond to LF offsets of $0.28^{+0.20}_{-0.20}$~dex overall, $0.35^{+0.23}_{-0.23}$~dex for the fainter bin, 
and $0.15^{+0.48}_{-0.52}$~dex for the brighter bin.

\section{Summary}

The key results from our work are as follows:
\begin{itemize}

\item{Using 26.48~deg$^2$ of deep ($\sim$26 magnitude) Subaru HSC $g$, $r$, $i$, $z$, and NB816 imaging of 
the NEP field, we identified 25 $z=5.7$ LAE candidates.}

\item{We spectroscopically observed 17 of the 18 most luminous ULLAE candidates with DEIMOS on Keck~II and
confirmed 9 as $z=5.7$ ULLAEs and 3 as $z=5.7$ sub-ULLAEs.}

\item{We constructed a $z=5.7$ ULLAE LF from our spectroscopic sample, which we corrected for incompleteness using a rigorous incompleteness simulation.}

\item{We compared our $z=5.7$ ULLAE LF and the reprocessed $z=6.6$ ULLAE LF from \cite{taylor20} with $z=5.7$ and $z=6.6$ LAE LFs from the literature. Our spectroscopic LFs can be viewed as ultraluminous extensions of the spectroscopic LFs from \cite{hu10}. There is good agreement where these LFs meet near the ultraluminous threshold.}

\item{We observed a $0.35^{+0.23}_{-0.23}$~dex drop from our 
$z=5.7$ LF to our $z=6.6$ LF for the $\log L(\rm \Lya)=43.50-43.75$~erg~s\per bin and 
a $0.15^{+0.48}_{-0.52}$~dex drop for the $\log L(\rm \Lya)=43.75-44$~erg~s\per bin.}

\end{itemize}

The present results together with the identification of double-peaked \Lya lines in the $z=6.6$ samples \citep{hu10,matthee18,songaila18,meyer21} suggest that we are seeing ionized bubbles around the most luminous LAEs.
Modeling of these structures can provide powerful constraints on the escape fractions and general
properties of these galaxies \citep{gronke20}.

\section{Acknowledgments} 
We thank the anonymous referee for constructive reports that helped us to improve the paper. 
We gratefully acknowledge support for this research from a Wisconsin Space Grant Consortium Graduate and Professional Research Fellowship (A.J.T.), a Sigma Xi Grant in Aid of Research (A.J.T.), NSF grants AST-1715145 (A.J.B) and AST-1716093 (E.M.H., A.S.), 
 the William F. Vilas Estate (A.J.B.), and a Kellett Mid-Career Award from the  
University of Wisconsin-Madison Office of the Vice Chancellor for Research and 
Graduate Education with funding from the Wisconsin Alumni Research Foundation (A.J.B.). 

This paper is based in part on data from the Subaru Telescope.
The Hyper Suprime-Cam (HSC) collaboration includes the astronomical communities 
of Japan and Taiwan, and Princeton University. The HSC instrumentation and software 
were developed by the National Astronomical Observatory of Japan (NAOJ), the Kavli 
Institute for the Physics and Mathematics of the Universe (Kavli IPMU), the University 
of Tokyo, the High Energy Accelerator Research Organization (KEK), the Academia 
Sinica Institute for Astronomy and Astrophysics in Taiwan (ASIAA), and Princeton 
University. Funding was contributed by the FIRST program from Japanese Cabinet 
Office, the Ministry of Education, Culture, Sports, Science and Technology (MEXT), 
the Japan Society for the Promotion of Science (JSPS), Japan Science and 
Technology Agency (JST), the Toray Science Foundation, NAOJ, Kavli IPMU, KEK, 
ASIAA, and Princeton University. 

This paper also makes use of data collected at the Subaru Telescope and retrieved 
from the HSC data archive system, which is operated by the Subaru Telescope and 
Astronomy Data Center at National Astronomical Observatory of Japan. 
Data analysis was in part carried out with the cooperation of Center for 
Computational Astrophysics, National Astronomical Observatory of Japan.

This paper is based in part on data collected from the Keck~II Telescope.
The W.~M.~Keck Observatory is operated as a scientific partnership among the 
California Institute of Technology, the University of California, and NASA, and was 
made possible by the generous financial support of the W.~M.~Keck Foundation.

This material is based upon work supported by NASA under Award No. RFP20\_9.0 issued through Wisconsin Space Grant Consortium. Any opinions, findings, and conclusions or recommendations expressed in this material are those of the authors and do not necessarily reflect the views of the National Aeronautics and Space Administration.

This research made use of \textit{Astropy}, a community-developed core Python package for Astronomy \citep{astropy:2013, astropy:2018}.

The authors wish to recognize and acknowledge the very significant cultural role and 
reverence that the summit of Maunakea has always had within the indigenous 
Hawaiian community. We are most fortunate to have the opportunity to conduct 
observations from this mountain.

\bibliography{Lyabib.bib}

\appendix
\section{1D Spectra}
\label{append:a}
\begin{figure*}[ht]
\includegraphics[width=\linewidth]{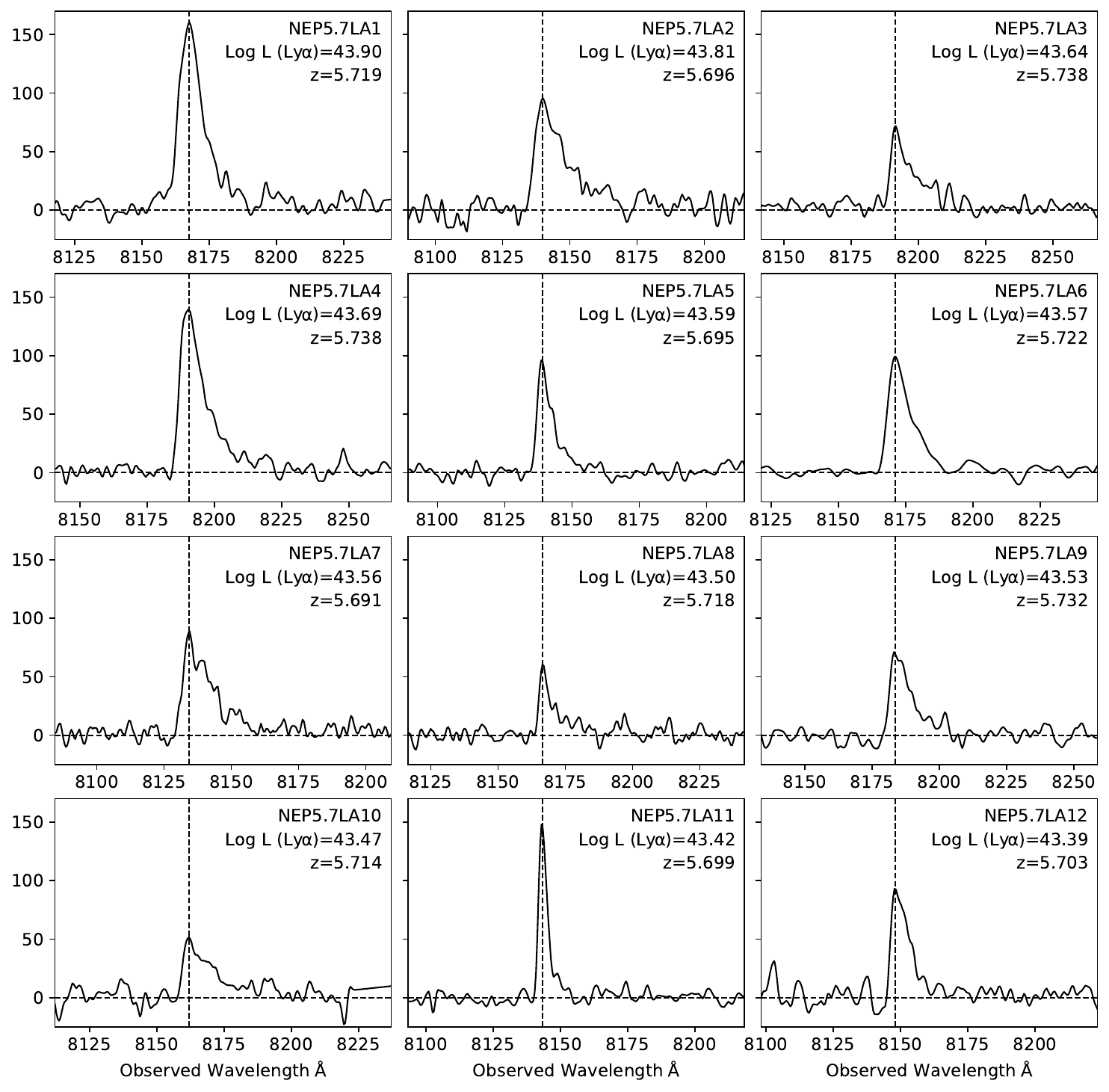}

\caption{1D spectra of the 12 confirmed $z=5.7$ ULLAEs and LAEs in the NEP field. 
The vertical scale is the flux in arbitrary units, but these are consistent between the spectra, so the normalizations can be directly compared. Note that the corresponding redshifts and narrowband calculated luminosities (in erg s\per) are given in the subplots.
}
\label{fig:1dspectra}
\end{figure*}

\begin{figure*}[ht]
\includegraphics[width=\linewidth]{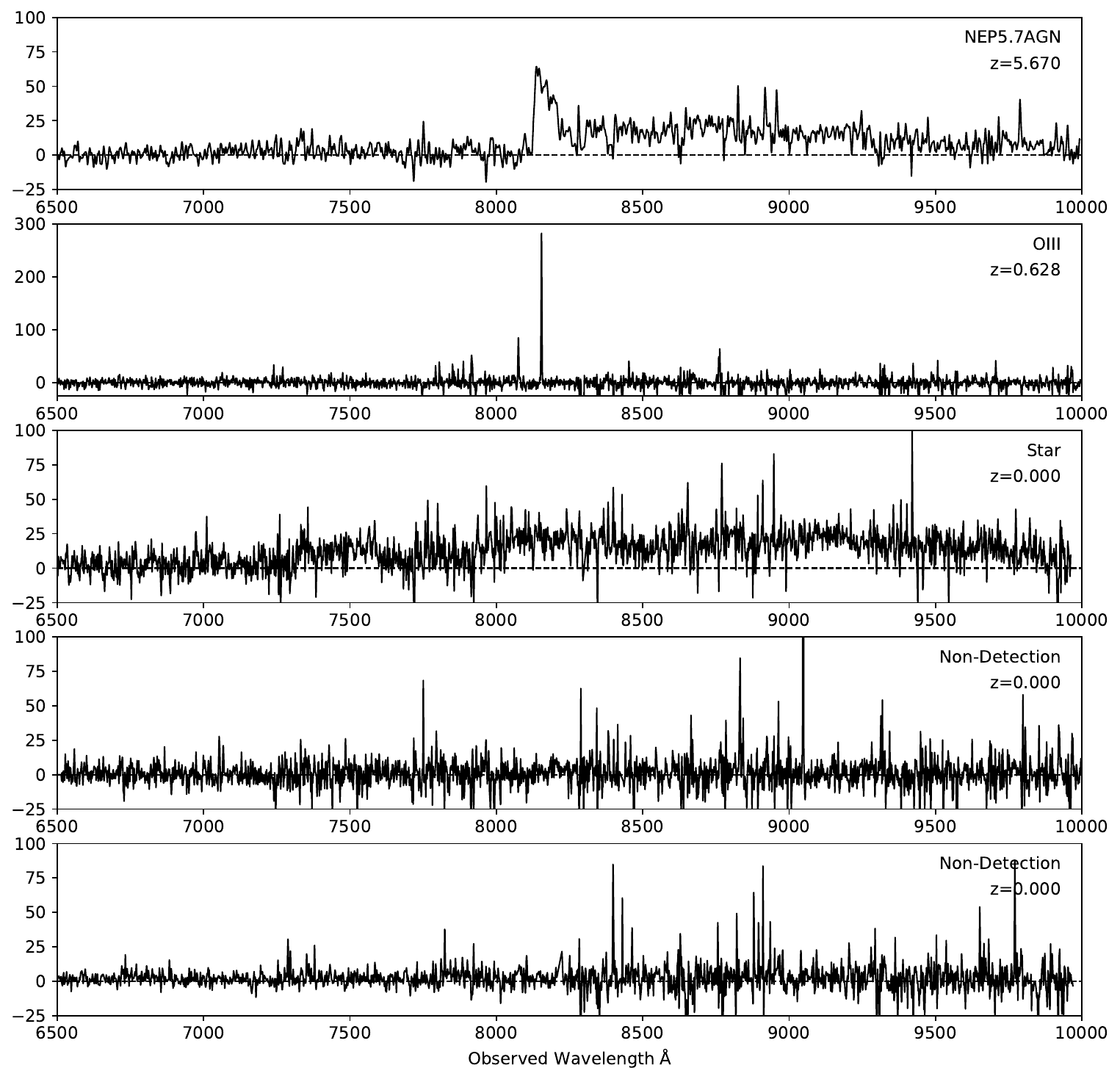}

\caption{1D spectra of the $z=5.670$ AGN, the $z=0.628$ [OIII] emitter, the red star, and the two spectroscopic non-detections in the NEP field. 
The vertical scale is the flux in arbitrary units, but these units are consistent between the spectra, so the normalizations can be directly compared. Note that the corresponding redshifts are given in the subplots. }
\label{fig:1dspectraother}
\end{figure*}

\end{document}